\documentclass[useAMS,usenatbib]{mn2e}
\usepackage{graphicx}
\usepackage{longtable}
\newcommand{\ms}{\mbox{ms$^{-1}~$}}
\newcommand{\apjs}{ApJS}
\newcommand{\aap}{A\&A}
\newcommand{\aapr}{A\&A Rev.}
\newcommand{\aj}{AJ}
\newcommand{\apj}{ApJ}
\newcommand{\mnras}{MNRAS}
\newcommand{\pasp}{PASP}
\newcommand{\nat}{Nature}
\newcommand{\actaa}{AcA}
\newcommand{\araa}{ARA\&A}

\title[Two super-Earths around Kapteyn's star]{
Two planets around Kapteyn's star : a cold and a temperate
super-Earth orbiting the nearest halo red-dwarf}

\author[Anglada-Escud\'e, Arriagada, Tuomi, Zechmeister, Jenkins, Ofir, Marvin et al.]{
G. Anglada-Escud\'e $^{1,2}$\thanks{E-mail:guillem.anglada@gmail.com},
P. Arriagada$^{3}$,
M. Tuomi$^{4,5}$,
M. Zechmeister$^{2}$,\newauthor
J. S. Jenkins$^{4}$,
A. Ofir$^{2}$,
S. Dreizler$^{2}$,
E. Gerlach$^{6}$,
C.~J. Marvin$^{2}$,
A. Reiners$^{2}$,\newauthor
S.~V. Jeffers$^{2}$,
R. Paul Butler$^{3}$,
S.~S. Vogt$^{7}$,
P.~J. Amado$^{8}$,
C. Rodr\'iguez-L\'opez$^{8}$,\newauthor
Z. M. Berdi\~nas$^{8}$,
J. Morin$^{9,2}$, 
J.~D. Crane$^{10}$,
S.~A. Shectman$^{10}$,
I.~B. Thompson$^{10}$,
\newauthor
M. D\'iaz$^{5}$,
E. Rivera$^{7}$,
L.~F.~Sarmiento$^{2}$,
H.~R.~A. Jones$^{5}$
\\
$^{1}$
School of Physics and Astronomy, Queen Mary, University of London,
327 Mile End Rd. London, United Kingdom\\
$^{2}$
Universit\"{a}t G\"{o}ttingen,
Institut f\"ur Astrophysik,
Friedrich-Hund-Platz 1,
37077 G\"{o}ttingen, Germany\\
$^{3}$
Carnegie Institution of Washington,
Dept. of Terrestrial Magnetism,
5241 Broad Branch Rd. NW, 20015,
Washington D.C., USA\\
$^{4}$
Departamento de Astronom\'ia, Universidad de Chile,
Camino El Observatorio 1515, Las Condes, Santiago, Chile,
Casilla 36-D.\\
$^{5}$
Centre for Astrophysics Research,
University of Hertfordshire,
College Lane, AL10 9AB, Hatfield, UK\\
$^{6}$
Institut f\"ur Planetare Geod\"asie
Technische Universit\"at Dresden
01062, Dresden, Germany\\
$^{7}$
UCO/Lick Observatory, University of California,
Santa Cruz, CA, 95064, USA\\
$^{8}$
Instituto de Astrof\'isica de Andaluc\'ia-CSIC,
Glorieta de la astronom\'i a S/N, 18008, Granada, Spain\\
$^{9}$
LUPM-UMR5299, CNRS \& Universit\'e Montpellier II, Place E. Bataillon,
Montpellier, F-34095, France\\
$^{10}$
Carnegie Institution of Washington, The Observatories, 
813 Santa Barbara Street, Pasadena, CA 91101-1292, USA
}
\begin{document}
\bibliographystyle{mn2e}

\date{Submitted April 25, 2014}

\pagerange{\pageref{firstpage}--\pageref{lastpage}} \pubyear{2014}

\maketitle

\label{firstpage}

\begin{abstract} 
Exoplanets of a few Earth masses can be now detected around nearby
low-mass stars using Doppler spectroscopy. In this paper, we 
investigate the radial velocity variations of Kapteyn's star, which is
both a sub-dwarf M-star and the nearest halo object to the Sun. The
observations comprise archival and new HARPS, HIRES and PFS Doppler
measurements. Two Doppler signals are detected at periods of 48 and
120 days using likelihood periodograms and a Bayesian analysis of
the data. Using the same techniques, the activity indicies and
archival ASAS-3 photometry show evidence for low-level activity
periodicities of the order of several hundred days. However, there
are no significant correlations with the radial velocity variations
on the same time-scales.  The inclusion of planetary Keplerian
signals in the model results in levels of correlated and excess
white noise that are remarkably low compared to younger G, K and M
dwarfs. We conclude that Kapteyn's star is most probably orbited by
two super-Earth mass planets, one of which is orbiting in its
circumstellar habitable zone, becoming the oldest potentially
habitable planet known to date. The presence and long-term survival
of a planetary system seems a remarkable feat given the peculiar
origin and kinematic history of Kapteyn's star. The detection of
super-Earth mass planets around halo stars provides important 
insights into planet-formation processes in the early days of 
the Milky Way.
\end{abstract}

\begin{keywords}
techniques: radial velocities -- stars: individual: Kapteyn's star, planetary systems
\end{keywords}

\section{Introduction}

Sub-\ms velocity precision can now be achieved for M-dwarfs resulting  from
stabilized spectrographs \citep[such as HARPS, ][]{mayor:2003} and specialized
spectral analysis techniques \citep{anglada:2012a}. This increased sensitivity
enables the detection of planets of a few-Earth masses orbiting in the star's
habitable zone.  Recently, statistical population analyses of the NASA/Kepler
mission \citep{dressing:2013}, and ground-based Doppler surveys
\citep{bonfils:2013, tuomi:2014:uves} suggests that every low-mass star has at
least one (or more) planet with an orbital period less than 50 days. Very small
planets in very short-period orbits have recently been discovered by the Kepler
mission \citep[e.g.~KOI-1843, ][]{ofir:2013}. We recently started the Cool Tiny
Beats survey to charaterise the Doppler variability over short time-scales and
search for such small planets in short period orbits around nearby low-mass
stars. The program uses the High Accuracy Radial velocity Planet Searcher
spectrograph (HARPS) installed at the 3.6m ESO telescope at La Silla/Chile and
its northern counterpart at the Telescopio Nationale Galileo/La Palma. 

\begin{figure*}
\center
\hspace{0.01in}
\includegraphics[width=0.40\textwidth,clip]{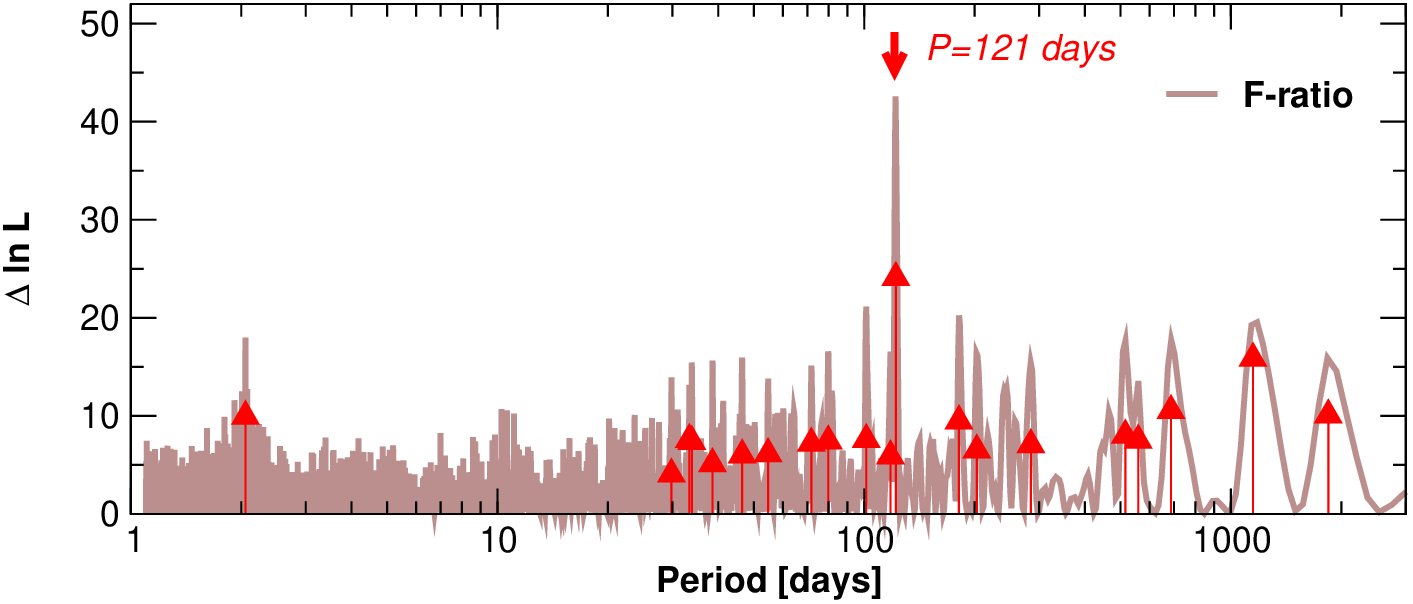}
\includegraphics[width=0.40\textwidth,clip]{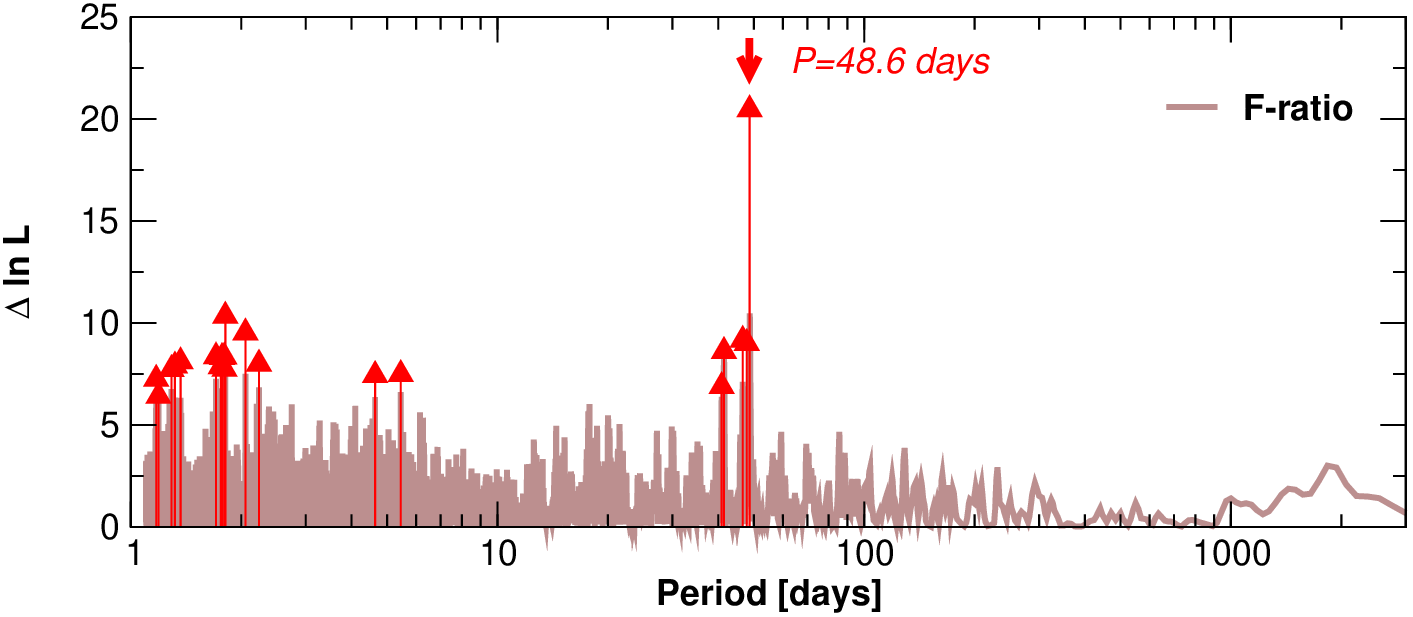}
\includegraphics[width=0.40\textwidth,clip]{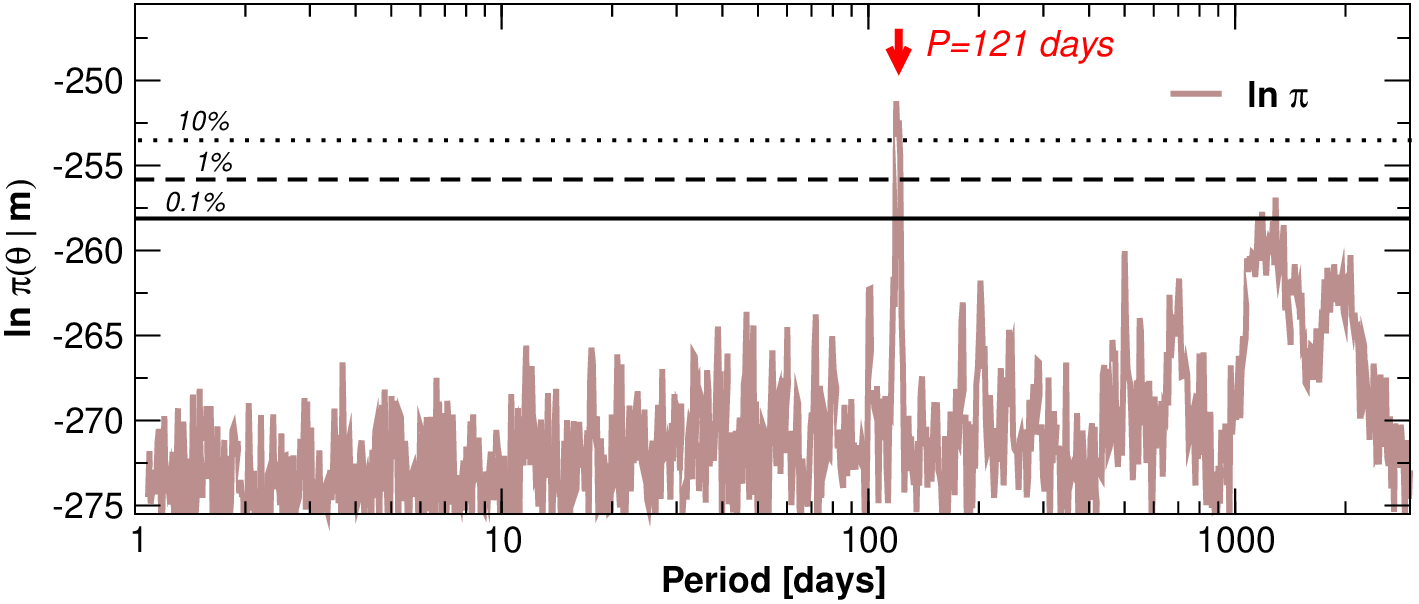}
\includegraphics[width=0.40\textwidth,clip]{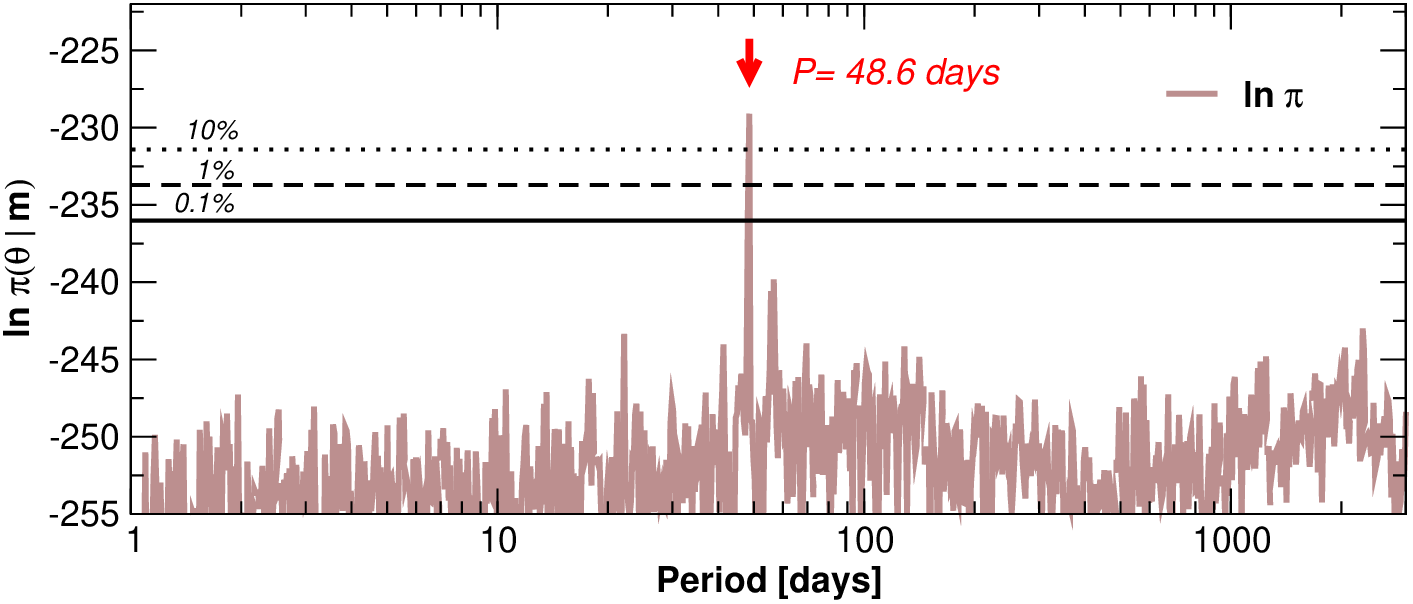}

\caption{Top panels illustrate the detection periodograms for a
model with one-planet (left), and the search for the second
signal (right).  Brown thick line is the F-ratio periodogram 
applied to residual data, it does not adjust for the 
extra-white noise as a free parameter and is only used to initialize 
the global maximum-likelihood searches. Red
triangles represent the top twenty likelihood maxima. Bottom panels 
show the Posterior probability contours as obtained by tempered MCMC
samplings using the DRAM algorithm (see text). Horizontal lines
indicate relative probability thresholds compared to the
maximum posterior. For perfectly known uncertainties and in
the absence of correlations, the F-ratio and the
log-likelihood ratio statistic should coincide. However, this
is hardly the case when dealing with Doppler data. Top panels
also illustrate the perils of working with residual data
statistics to assess significance of
signals. That is, the F-ratio is over-optimistic in the
significance of the first candidate (left panel, 121
days brown peak is much higher), while the significance of the
second signal at 48.6 days (right panel) would be greatly
underestimated.} 
\label{fig:periodograms} 
\end{figure*}

\begin{table*}
\caption{Spectroscopic measurements.
Median value and a perspective acceleration were subtracted to each RVs set 
(Ins. 1 is HARPS, 2 is HIRES, 3 is PFS). 
FWHM, BIS, and S-index are provided for HARPS only. 
Uncertainty in the FWHM is $2.5\times\sigma_{\rm BIS}$.}
\label{tab:data}
\begin{center}
\tiny
\begin{tabular}{lllllllllll}
\hline \hline
JD            &  RV          &  $\sigma_{\rm RV}$  &  Ins  & BIS           & $\sigma_{\rm BIS}$ & FWHM           & S-index   & $\sigma_{\rm S-index}$\\
$[$days$]$    &  $[$\ms$]$   & $[$ms$^{-1}]$       &       & $[$ms$^{-1}]$ & $[$ms$^{-1}]$      & $[$kms$^{-1}]$ & $[$-$]$   & $[$-$]$               \\
\hline
2452985.74111 &  3.11        &   0.80              &   1   &  -4.64        &  0.86              &   3.21561      &  0.2966   &  0.0042               \\
2452996.76321 &  3.00        &   0.27              &   1   &  -9.75        &  0.56              &   3.19206      &  0.2699   &  0.0033               \\
2453337.80096 & -2.65        &   0.89              &   1   &  -8.17        &  0.76              &   3.18036      &  0.1859   &  0.0036               \\
2453668.83537 & -2.57        &   0.52              &   1   & -10.27        &  0.49              &   3.21216      &  0.2678   &  0.0026               \\
\ldots\\
\hline \hline
\end{tabular}
\end{center}
\end{table*}

\section{Kapteyn's star}
At only 3.91 pc, Kapteyn's star (or GJ 191, HD 33793) is the
closest halo star to the Sun \citep{hipparcos}. It is subluminous
with respect to main sequence stars of the same spectral type and
was spectroscopically classified as an M1.0 sub-dwarf by
\citet{gizis:1997}. A radius of 0.291$\pm$0.025 R$_\odot$ was
directly measured by \citet{segransan:2003} using interferometry.
This combined with an estimate of its total luminosity was then
used to derive an effective temperature of 3570$\pm$156 K and its
mass was estimated to be 0.281$\pm$0.014 M$_\odot$.
\citet{woolf:2005} determined a metallicity of $[M/H]$ = -0.86
which coincides with several later estimates within 0.05 dex. We
independently obtained estimates of its metallicity and temperature
by comparing its observed colors (B, V, J, H, and K) to
synthetically generated ones from the PHOENIX library
\citep{husser:2013}, obtaining compatible values of ${\rm[Fe/H]}=-0.89$
and $3550\pm50$ K. Only an upper limit has been measured for its
projected rotation velocity \citep[$v\sin i < 3 { \rm km s}^{-1}$, ][]{browning:2010}, 
and its X-ray luminosity has been measured
to be comparable to other multi-planet host M-dwarfs such as GJ~876
and GJ~581 \citep{walkowicz:2008}. Therefore, planets of a few 
Earth-masses orbiting such an inactive M-star should be detectable
using Doppler spectroscopy \citep{barnes:2011}.
A precise age estimation of the star cannot be obtained from 
models, as they change very little for M~$<$~0.6~M$_\odot$ in 
the range of ages between 0.4 and 15~Gyr 
\citep{baraffe:1997, segransan:2003}. The low metallicity and
halo kinematics suggest an ancient origin, which is consistent
with its low-activity and slow rotation.

\begin{figure}
\center
\includegraphics[width=0.30\textwidth,clip]{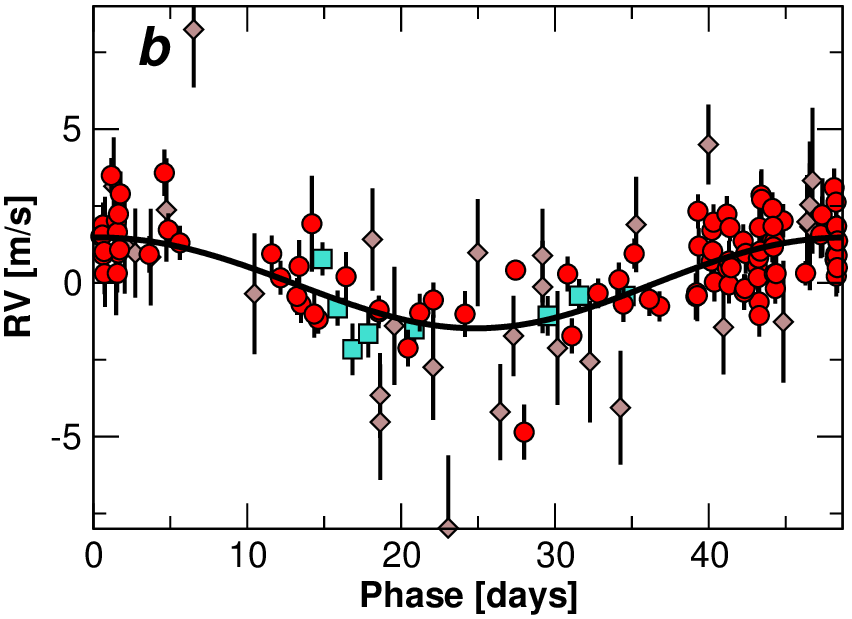}
\includegraphics[width=0.30\textwidth,clip]{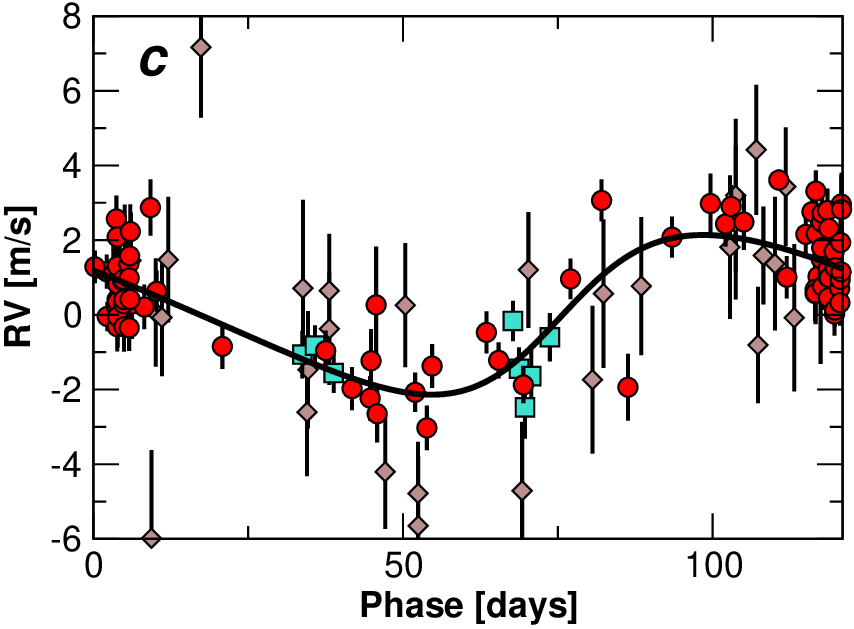}
\caption{Phase folded Doppler curves to the reported signals
with the other signal removed (HARPS are red circles, HIRES 
are brown diamonds and PFS are blue squares). The maximum
likelihood solution is depicted as a black line.}
\label{fig:phased}
\end{figure}

\section{Observations}
The observations comprise both new and archival data from HARPS,
HIRES and PFS spectrometers. HARPS is a stabilized high-resolution
spectrometer (resolving power $\sim$ 110 000) covering from 380 to
680 nm. The HARPS data we use come from two programs; the Cool Tiny
Beats survey and archival data from the first HARPS-GTO survey. The
Cool Tiny Beats data were obtained in two 12 night runs in May 2013
(11 spectra) and in Dec 2013 (55 spectra). The HARPS-GTO data
contain 30 spectra taken between 2003 and 2009.
\citet{bonfils:2013} reported some possible signals on Kapteyn's
star, but no detection claim could be made at the time. Doppler
measurements were obtained using the least-squares
template-matching approach as implemented by the HARPS-TERRA
software \citep{anglada:2012a}. For each spectrum, we also use two
measurements of the symmetry of the mean-line profile as provided
by the HARPS-Data Reduction Software. These are the bisector
span (BIS) and the full-width-at-half-maximum (FWHM) of the
cross-correlation function. BIS and FWHM are known to correlate
with activity induced features that can cause spurious Doppler
signals. For example, changes in the line-symmetry caused by
co-rotating dark and magnetic spots induce changes in the symmetry
of the lines that produce spurious Doppler shifts
\citep[e.g.,][]{reiners:2013}. The variability in the Ca II H+K
emission lines was also measured by HARPS-TERRA through the
S-index. The variability in the S-index is also known to correlate with
magnetic activity of the star \citep{dasilva:2012}, and localized
active regions such as spots. Any Doppler signal with a period
equivalent to variability detected in the BIS, FWHM and S-index is
likely to be spurious.

We include 30 Doppler measurements spanning between 1999 and 2008
using the HIRES spectrometer at Keck \citep{vogt:1994}. These
measurements were obtained using the Iodine cell technique as
described in \citet{butler:1996}. A new HIRES stellar template was
generated by deconvolving a high signal-to-noise
iodine-free spectrum of the star applying Maximum Likelihood
deconvolution with Boosting \citep{morhac:2009}. The star
is never at an altitude greater than 26 deg from Hawaii, which
could explain the lower accuracy compared to the HARPS
measurements. The long baseline of HIRES puts
strong constraints on the long-period variability and mitigates
alias ambiguities in the HARPS data. We also included
8 new Doppler measurements obtained with PFS
\citet{crane:2010} at Magellan/Las Campanas observatory, also using
the Iodine cell technique. While the statistical contribution of
PFS is small, the significance of the preferred solution increases
thus providing further support to the signals. All spectroscopic 
measurements used are given in table \ref{tab:data}.

\section{Signal detection methods}

The initial signal detection is performed using log-likelihood
periodograms \citep{baluev:2009, anglada:2013} and tempered Markov
Chain samplings using the delayed-rejection adaptive-Metropolis
method \citep[or DRAM, ][]{haario:2006} as implemented in
\citet{tuomi:corot:2014}. These methods produce maps of the maximum
likelihood and posterior densities as a function of the period
being investigated. As in classic periodograms, preferred periods
will appear as peaks (local probability maxima). The significance
is then assessed using the Bayesian criteria described
in \citet{tuomi:2014:uves}. We say that a Doppler signal is
detected if 1) its period is well-constrained, 2) its amplitude is
statistically significantly different from zero, and 3) inclusion
of the signal in the model increases the Bayesian evidence by a
factor of $10^4$, which is more conservative than usual because
important effects might be missing in the model, especially when
combining data from different instruments. The model of the
observations is fully encoded in the definition of the likelihood
function. In addition to the usual Keplerian parameters, our model
includes three nuisance parameters for each instrument (INS): a
constant offset $\gamma_{\rm INS}$, an extra white-noise term
$\sigma_{\rm INS}$ and a Moving Average coefficient $\phi_{\rm
INS}$ that quantifies the amount of correlation between consecutive
measurements. $\phi_{\rm INS}$ is bound between $+1$ and $-1$
(anticorrelation) and a value consistent with zero implies no
significant correlation. Our model also assumes an exponential
decay of the correlation that depends on a characteristic
time-scale $\tau_{\rm INS}$. Adding it as a free parameter led to
overparameterization, so a fixed value of $\tau_{\rm INS}=4$ days 
was set by default. Linear trends caused by the presence
of planets with periods much longer than the time-baseline
is parameterized as $\dot{\gamma} (t_i-t_0)$, where $\dot{\gamma}$
is also a free parameter of the Doppler model and $t_0$ is
an arbitrary reference epoch (see Table \ref{tab:parameters}).
The log-likelihood periodograms use the same likelihood
model but do not include correlations terms. False alarm
probability assessments using likelihood periodograms \citep[or
FAP, see][]{baluev:2008} are also computed to double-check the
significance of each detection.

\section{Analysis of the time-series}

The search for signals in the Doppler time-series is summarized in
Figure \ref{fig:periodograms}. Two very significant periods are
detected at 121 and 48.6 days. The Bayesian evidence ratios are
$3.2\times10^9$ for the one-planet model against the no-planet
model, and $3.1\times10^9$ for the two-planet solution against the
one-planet model. These numbers are well in excess of the 10$^4$
threshold indicating very confident detections. The FAPs as derived
from the log-likelihood periodograms are $1.5\times10^{-6}$ and
$8.9\times10^{-5}$ respectively, which are also very small further
supporting the detections (acceptable detection thresholds using 
periodogram methods are typically between 1\% and 0.1\%). The 
support of the signals by the three data sets is illustrated in the
phase-folded curves in Figure~\ref{fig:phased}. The uneven
sampling can produce biases in the significance estimates. To
investigate this, we also performed the same analysis using nightly
averaged measurements. The same two signals are detected well above
our significance thresholds, thus providing further confidence in
the detections. As for other stars, further follow-up over the next
few years is desirable to confirm that both low-amplitude signals
remain coherent over time.

We investigated linear correlations with activity indices by
including them in the likelihood model of the HARPS data. While we
could still easily detect the planetary signals, these models had
lower probabilities due to over-parameterization and are not
included in the final solution. We also verified whether if any
indicator of activity showed variability in similar time-scales as
the Doppler data. While the BIS does not show any hint of temporal
coherence, the FWHM and S-index do share similar periodogram
structures with tentative peaks between 140 to 2000 days. However,
neither had FAPs that were lower than 5\%, indicating little
significance. 

\begin{table}
\caption{The Keplerian solution of the combined radial velocities 
presented as median of the posterior estimates and corresponding 
68\% credibility intervals. The uncertainties of semi-major axes 
and minimum masses include uncertainties in the stellar mass. 
Reference epoch $t_0$ for computation of $M_0$ (mean anomaly) and 
$\lambda_0$ (mean longitude) is assumed to be JD=2452985.74111 days.}
\label{tab:parameters}
\begin{center}
\begin{tabular}{lrr}
\hline \hline
Model parameter & Kapteyn b & Kapteyn c\\
\hline
$P$ [day]      & $ 48.616^{+0.036}_{-0.032}$  & $121.54^{+0.25}_{-0.25}$  \\
$K$ [\ms]      & $  2.25^{+0.31}_{-0.31}$     & $  2.27^{+0.28}_{-0.26}$  \\
$e$            & $  0.21^{+0.11}_{-0.10}$     & $  0.23^{+0.10}_{-0.12}$  \\
$\omega$ [deg] & $ 80.4^{+30.7}_{-28.9}$      & $  3.9^{+28.9}_{-33.2}$   \\
$M_{0}$  [deg] & $269.6^{+38.3}_{-32.2}$      & $357.6^{+32.3}_{-27.9}$   \\
$\dot{\gamma}$ [\ms yr$^{-1}$] & $-0.181^{+0.088}_{-0.086}$ \\
$\sigma_{\rm HARPS}$ [\ms] & $0.65^{+0.10}_{-0.10}$     \\
$\sigma_{\rm HIRES}$ [\ms] & $1.17^{+0.51}_{-0.47}$     \\
$\sigma_{\rm PFS}$ [\ms]   & $0.32^{+0.69}_{-0.43}$     \\
$\phi_{\rm HARPS}$ & $0.03^{+0.10}_{-0.08}$ \\
$\phi_{\rm HIRES}$ & $0.40^{+0.50}_{-0.50}$   \\
$\phi_{\rm PFS}$   & $0.10^{+0.30}_{-0.30}$ \\
\\
\multicolumn{2}{l}{Derived quantities}\\
\hline
$\lambda_0 = \omega+M_0$ [deg] & $350^{+19.1}_{-18.7}$    & $1.6^{+20.5}_{-20.1}$ \\
$m_{p} \sin i$ [M$_{\oplus}$]  & 4.8$^{+0.9}_{-1.0}$       &  7.0$ ^{+1.2}_{-1.0}$   \\
$a$ [AU]                       & 0.168$^{+0.006}_{-0.008}$ &  0.311$^{0.038}_{0.014}$ \\
S/S$_\oplus$                   & 40\%                      & 12\%   \\
HZ-range [AU]                 & $\sim$0.126-0.236  \\
P$_c$/P$_b$                   & $2.496^{+0.021}_{-0.029}$ \\
\hline \hline
\end{tabular}
\end{center}
\end{table}

We also analysed V-band photometry obtained from the latest release
of the ASAS program \citep{asas}\footnote{ASAS-3 release
\texttt{http://www.astrouw.edu.pl/asas/?page=catalogues}}. A
standard deviation computed using the central 80\% percentile was
used to remove 4-$\sigma$ outliers resulting in 503 epochs spanning
from Dec 2000 to Sep 2007. A log-likelihood periodogram analysis
identified two significant signals at 340 and 1100 days, both with
amplitudes of $\sim$ 5 mmag. The signal at 1100 days is compatible
with other reports of periodicities in activity indices of M-dwarfs
\citep{dasilva:2012}, and the 340-day variability might be caused
by seasonal systematic errors, or by a long rotation period. The
periodogram structure of the photometry resembles those of the FWHM
and the S-index, further supporting potential low-level activity
changes happening at timescales of several hundred days. Given that
the activity and Doppler signals appear uncorrelated, we conclude
that the simplest interpretation of the Doppler data is the
presence of two planets (see Table \ref{tab:parameters}). With a
period of 48.6 days, Kapteyn~b lies well within the liquid water
habitable zone of the star \citep{Kopparapu13}. Assuming rocky
composition, its properties and possible climates should be similar
to those discussed for GJ~667Ce \citep{anglada:2013}. Kapteyn~c
receives $\sim$ 10\% of Earth's irradiance, implying that the
possibility of it being able to support liquid water on its surface
is less probable. 

We used Optimal BLS \citep{ofir:2014} to search for possible transit
events in the ASAS photometry. No trasits were found when searching
for a signal in the range of parameters compatible with the Doppler
solution. Further observations at the predicted transit windows are
needed to put meaningful constraints on possible transit signals.

\section{Discussion}

The age of Kapteyn star can be inferred from its membership to the
Galactic halo and peculiar element abundances \citep{kotoneva:2005}.
The current hierarchical Milky Way formation scenario suggests that
streams of halo stars were originated  as tidal debris from satellite
dwarf galaxies being engulfed by the early Milky Way \citep[][ and
references therein]{klement:2010}. This scenario is supported by the
age estimations of the stars in the inner halo \citep[$\sim$10--12 Gyr
,][]{jofre:2011}, and globular clusters. The work by Eggen and
collaborators \citep[see, ][as review summary]{eggen:1996}
established  the existence of such high-velocity, metal-poor moving
groups in the solar neighborhood. Kapteyn's star is the prototype
member of one of these groups, which has been recently investigated by
\citet{wylie-deBoer:2010} among others. Stars in Kapteyn's group share
a retrograde velocity of rotation around the Galactic centre of about
$-290$ k\ms \citep{eggen:1996}. Spectroscopic observation of 16 members
has shown that the group likely originated from the same progenitor
structure as the peculiar globular cluster $\omega$ Centauri, but not
from within the cluster itself \citep{kotoneva:2005,
wylie-deBoer:2010}. The origin of Kapteyn's star within a merging
dwarf-galaxy sets its most likely age around the young halo's one ($>
10$ Gyr) and should not be older than 13.7 Gyrs, which is the current
estimate of the age of the universe \citep{planck}. 

Numerical integration of several possible orbits over 10$^4$yr
to 1~Myr using \textit{Mercury 6} \citep{chambers:1999} shows
very small changes in the orbital parameters. Despite the
the periods for the two planet candidates are compatible with a 5:2
period commensurability (see Table \ref{tab:parameters}), the analysis
of the resonant angles for the best fit solution did not support 
the presence of dynamical mean-motion resonances, which existence
also depend on a number of other properties such total planet masses
and mutual inclinations. We found, however, that the difference
in periastron  angles $\Delta \omega=\omega_b-\omega_c$
librates around 180 deg; which corresponds to a long-term stable 
configuration called apsidal locking. This is sufficient 
to show that the proposed system is compatible with long-term 
physically-viable solutions. It has been suggested that two-planet 
systems that underwent weak dissipation (slow migration) should always
end-up in absidal locking \citep{michtchenko:2011}, which has
consequences on the likely planet-formation scenario. A
more thorough analysis to properly quantify the significance of
the absidal locking and possible mean-motion ressonances
requires a much more extended discussion
\citep[e.g.,][]{anglada:2013} which is beyond the scope of this
paper. 

The detection of two super-Earths here is consistent with the
idea that low-metallicity stars are more prone to the formation of
low-mass planets rather than gas giants
\citep{udry:2007,buchhave:2012}. This is further supported by a
significant paucity of lowest-mass planets in Doppler searches of
metal-rich stars \citet{jenkins:2013}, and non-comfirmation of
previous claims of gas giants around extremely metal-poor stars
\citep{desidera:2013, jones:2014}. These observational findings are
compatible with recent population synthesis experiments such as those
in \citet{mordasini:2012}.

Once the planets signals are included in the Doppler model, the RV
residuals variability are reduced to instrumental noise (i.e.~extra
jitter term is compatible with reported stability of HARPS, and
correlation coefficients compatible with 0). This indicates that the
star is very Doppler stable, possibly more stable than the instruments
themselves. This indeed would be expected because pulsation and
convective motions are thought to be much smaller in inactive low-mass
stars than in earlier types. At the likely age of the system, most G
and K dwarfs are evolving away from the main sequence into giants,
which makes the Doppler detection of small planets unfeasible due to
increased activity levels \citep[e.g.,~][]{nowak:2013}. As a result,
original architectures of the first planetary systems can only be
explored by observing venerable low-mass stars which are still on the
main-sequence such as Kapteyn's star.

\noindent \textbf{Acknowledgments.} We thank R.P.Nelson,
J.Chanam\'e and R.~Baluev (referee) for constructive comments
and discussions. We acknowledge funding from : DFG/Germany
through CRC-963 (CJM); DFG/Germany 1664/9-1 (AR); Alexander Von
Humboldt Foundation/Germany (JM); ERC-FP7/EU grant number 27347
(MZ), AYA2011-30147-C03-01 by MINECO/Spain, FEDER funds/EU, and
2011 FQM 7363 of Junta de Andaluc\'ia/Spain (CR-L and PJA);
JAE-Doc program (CR-L); FPI BES-2011-049647 MINECO/Spain (ZMB);
CATA (PB06, CONICYT)/Chile (JSJ); CONICYT-PFCHA/Doctorado
Nacional/Chile (MD);NSF/USA grants AST-0307493 and
AST-0908870 (SSV); and NASA grant NNX13AF60G S02 (RPB). Based
on observations made with ESO Telescopes under programme ID
191.C-0505 and ESO's Science Archive Facility (req. number
GANGLADA95087). We acknowledges the effort of the HARPS team
program obtaining data within program 072.C-0488. Some data
were obtained at the W.M.~Keck Obs. made possible by the
support of the W.M. Keck Foundation and operated among Caltech,
Univ. of California and NASA. The authors are most fortunate to
conduct observations from the sacred mountain of Mauna Kea.
This study uses data obtained at Magellan, operated by the
Carnegie Inst., Harvard Univ., Univ. of Michigan, Univ. of
Arizona, and the Massachusetts Inst. of Technology.


\clearpage
\onecolumn
\appendix
\section{On-line table}

\begin{verbatim}
Spectroscopic measurements on Kapteyn's star (Anglada-Escude+, 2014)
================================================================================

Description:
Time-series of spectroscopic measurements used in the paper.
Median value and a perspective acceleration were subtracted to 
each RVs set  (Ins. 1 is HARPS, 2 is HIRES, 3 is PFS). 
Measurements of the FWHM, BIS of the cross-correlation
profiles and measurements of the S-index are provided for
HARPS data only. Uncertainty in the FWHM is 2.5 times
the uncertainty in BIS. Check 2012ApJS..200...15A, 
for more detailed definitions of the measurements and their uses.

--------------------------------------------------------------------------------
   Bytes Format Units   Label    Explanations
--------------------------------------------------------------------------------
   1- 13  F13.5 d       JD       Barycentric Julian date
  15- 19  F5.2  m/s     RVel     Radial velocity
  21- 24  F4.2  m/s     eRVel    Uncertainy in RV
  26      I1    ---     Ins      Instrument used
  28- 33  F6.2  m/s     BIS      Bisector span of the CCF
  35- 38  F4.2  m/s     eBIS     Uncertainty in BIS
  40- 46  F7.5  km/s    FWHM     FWHM of the CCF (1)
  48- 53  F6.4  ---     Sidx     CaII H+K S-index in the Mount Wilson system
  55- 60  F6.4  ---     eSidx    Uncertainty in S-index
--------------------------------------------------------------------------------

Note (1): Uncertainty in FWHM is 2.5x eBIS

================================================================================
2452985.74111  3.11 0.80 1  -4.64 0.86 3.21561 0.2966 0.0042
2452996.76321  3.00 0.27 1  -9.75 0.56 3.19206 0.2699 0.0033
2453337.80096 -2.65 0.89 1  -8.17 0.76 3.18036 0.1859 0.0036
2453668.83537 -2.57 0.52 1 -10.27 0.49 3.21216 0.2678 0.0026
2453670.73947 -3.76 0.59 1 -13.27 0.41 3.20937 0.2636 0.0022
2454141.57444  0.56 0.53 1 -13.62 0.37 3.20616 0.2968 0.0021
2454167.53053 -1.22 0.56 1 -14.86 0.38 3.20717 0.2616 0.0020
2454169.50521 -1.66 0.48 1 -13.34 0.34 3.19959 0.2578 0.0018
2454173.50080 -1.57 0.48 1 -13.76 0.36 3.19892 0.2540 0.0019
2454197.51207  2.47 0.54 1 -16.21 0.43 3.20409 0.2470 0.0022
2454229.46873  2.59 0.73 1 -13.00 0.56 3.20358 0.2926 0.0034
2454342.86258  2.61 0.59 1 -13.37 0.39 3.20687 0.2632 0.0020
2454345.87966  0.29 0.45 1 -12.34 0.33 3.20096 0.2549 0.0017
2454349.83417  0.35 0.45 1 -17.60 0.35 3.20115 0.2346 0.0017
2454424.73161  2.40 0.54 1 -14.00 0.37 3.19613 0.2305 0.0018
2454429.71897  4.68 0.56 1 -12.71 0.39 3.19832 0.2390 0.0020
2454449.76051  1.27 0.61 1 -13.88 0.45 3.18970 0.2318 0.0023
2454450.64700  1.67 0.57 1 -13.85 0.40 3.19968 0.2346 0.0020
2454452.70683  1.15 0.74 1 -17.21 0.53 3.19416 0.2540 0.0028
2454459.64359  0.16 0.56 1 -15.85 0.35 3.20016 0.2377 0.0017
2454462.70572  1.82 0.55 1 -10.81 0.35 3.20233 0.2457 0.0018
2454463.69854  2.61 0.49 1 -12.56 0.43 3.19982 0.2341 0.0022
2454464.68243  1.07 0.53 1 -12.38 0.42 3.20015 0.2452 0.0021
2454733.89725 -0.93 0.60 1 -15.14 0.36 3.20390 0.2536 0.0018
2454754.83883 -2.37 0.57 1 -13.84 0.41 3.20737 0.2538 0.0021
2454767.75805  0.08 0.58 1 -14.73 0.43 3.20601 0.2393 0.0021
2454879.55605 -2.33 0.59 1 -15.14 0.43 3.19513 0.4865 0.0032
2454879.69685 -1.36 0.85 1 -14.42 0.68 3.19309 0.2085 0.0034
2454880.52165 -0.01 1.56 1  -0.65 1.28 3.19636 0.2472 0.0053
2454880.67387 -2.96 0.76 1 -13.39 0.57 3.19398 0.2071 0.0028
2456417.46007 -0.10 0.60 1 -14.56 0.68 3.20851 0.2682 0.0044
2456417.46745 -0.54 0.68 1  -7.42 0.75 3.20894 0.2566 0.0047
2456418.46471  1.80 0.43 1 -16.75 0.55 3.20068 0.2761 0.0040
2456420.47541  0.63 0.39 1 -15.72 0.47 3.19810 0.3089 0.0037
2456421.46438  1.82 0.76 1 -18.81 0.95 3.20174 0.2652 0.0056
2456422.45767  2.00 0.72 1 -13.54 0.71 3.19709 0.2700 0.0049
2456423.46236  2.00 0.73 1 -11.24 0.89 3.19925 0.1930 0.0048
2456424.44602  1.03 0.93 1  -7.98 0.95 3.20863 0.4363 0.0069
2456426.46387  0.84 0.59 1 -13.18 0.68 3.19599 0.2540 0.0046
2456427.44860  3.42 0.75 1 -14.61 0.78 3.20405 0.2690 0.0050
2456428.46156  1.08 0.55 1 -10.70 0.68 3.20416 0.4196 0.0058
2456656.56057  0.24 0.75 1 -14.31 0.68 3.21574 0.2927 0.0042
2456656.59693  0.36 0.60 1 -11.10 0.51 3.20988 0.2697 0.0033
2456656.64552  0.26 0.83 1 -22.11 0.72 3.21746 0.2505 0.0041
2456656.69364  3.00 0.55 1 -12.33 0.55 3.20616 0.2597 0.0037
2456656.74624  1.87 0.58 1 -15.74 0.55 3.20590 0.2707 0.0037
2456657.55783  1.33 0.75 1 -16.59 0.62 3.21076 0.2827 0.0040
2456657.59109  2.32 0.53 1 -13.13 0.58 3.20375 0.2525 0.0037
2456657.64325  1.65 0.64 1 -11.93 0.58 3.21207 0.2984 0.0038
2456657.70141  2.59 0.57 1 -19.43 0.52 3.19999 0.2884 0.0040
2456657.75751  0.63 0.62 1 -19.62 0.50 3.21424 0.2675 0.0032
2456658.57385  2.80 0.58 1 -15.94 0.59 3.20647 0.2747 0.0038
2456658.60490  1.03 0.51 1 -13.50 0.57 3.21336 0.2671 0.0036
2456658.70151  1.26 0.65 1 -18.40 0.51 3.22847 0.2596 0.0029
2456658.70928  0.49 0.59 1 -12.00 0.50 3.23253 0.2737 0.0029
2456658.76592  2.35 0.43 1 -16.13 0.46 3.21731 0.2722 0.0031
2456658.84514  1.04 0.55 1 -14.10 0.49 3.19162 0.2733 0.0037
2456659.63210  1.86 0.53 1 -17.67 0.47 3.21368 0.2745 0.0030
2456659.69120  0.19 0.57 1 -11.60 0.48 3.20922 0.2819 0.0034
2456659.74851  0.31 0.45 1 -15.85 0.49 3.20449 0.2833 0.0036
2456659.79823  1.45 0.66 1 -10.39 0.54 3.19687 0.2690 0.0040
2456660.53935  0.99 0.74 1 -13.38 0.75 3.20692 0.2893 0.0046
2456660.59008  0.61 0.61 1 -16.00 0.59 3.20940 0.2862 0.0039
2456660.62526  1.24 0.52 1 -14.84 0.51 3.21176 0.2766 0.0033
2456660.68361  2.24 0.61 1 -15.18 0.55 3.20553 0.2735 0.0038
2456660.74382  1.46 0.63 1 -15.76 0.66 3.21245 0.2705 0.0042
2456660.79408  3.28 0.76 1 -16.45 0.83 3.20489 0.2632 0.0051
2456660.82165  3.14 0.98 1 -10.63 1.01 3.20620 0.3202 0.0065
2456661.53634  2.80 0.51 1 -18.11 0.54 3.21203 0.2905 0.0035
2456661.56643  2.21 0.55 1 -12.63 0.56 3.20931 0.2886 0.0037
2456661.59856  1.54 0.67 1 -13.82 0.54 3.20993 0.2781 0.0035
2456661.65691  0.66 0.52 1 -11.49 0.57 3.21339 0.2699 0.0038
2456661.71555  0.18 0.48 1 -15.70 0.56 3.21309 0.2501 0.0036
2456661.77193  0.66 0.81 1  -9.30 0.68 3.20710 0.2444 0.0042
2456665.54012  3.21 0.62 1 -15.37 0.55 3.21136 0.2941 0.0035
2456665.61482  0.85 0.41 1 -13.40 0.52 3.21657 0.3057 0.0035
2456665.63272  0.89 0.86 1 -14.43 0.66 3.21348 0.3073 0.0046
2456665.64914  1.06 0.59 1 -15.94 0.57 3.21032 0.2836 0.0040
2456665.66524  2.73 0.69 1 -15.27 0.66 3.21198 0.2955 0.0046
2456665.68169  0.33 0.65 1 -17.38 0.60 3.21303 0.2737 0.0040
2456665.69927  1.95 0.59 1 -15.54 0.61 3.21404 0.2957 0.0043
2456665.73292  1.00 0.71 1 -12.16 0.64 3.20612 0.2586 0.0042
2456665.74854  0.60 0.82 1 -17.69 0.75 3.20405 0.2949 0.0053
2456665.76395  1.46 0.83 1 -19.09 0.82 3.20640 0.2686 0.0054
2456666.53144  1.56 0.53 1 -22.85 0.69 3.21604 0.3345 0.0049
2456666.62771  1.60 0.59 1 -18.01 0.59 3.22144 0.2866 0.0039
2456666.67780  0.95 0.64 1 -14.63 0.63 3.21306 0.2783 0.0043
2456666.74492  1.06 0.62 1 -15.17 0.59 3.20733 0.2932 0.0045
2456666.76073  0.33 0.66 1 -15.40 0.66 3.21056 0.2877 0.0047
2456667.54294  2.00 0.73 1 -16.22 0.82 3.22648 0.3151 0.0048
2456667.59735  0.29 0.61 1 -14.08 0.62 3.21076 0.2695 0.0040
2456667.60350  1.62 0.56 1 -13.13 0.62 3.21454 0.3004 0.0042
2456667.67412  2.21 0.79 1 -16.97 0.72 3.21627 0.3009 0.0049
2456667.74602  1.04 0.57 1 -14.67 0.64 3.20707 0.3147 0.0048
2456667.79693  2.86 0.72 1  -9.98 0.78 3.21002 0.2869 0.0052
2451170.92507  5.00 1.59 2   0.00 0.00 0.00000 0.0000 0.0000
2451580.81122 -4.06 1.71 2   0.00 0.00 0.00000 0.0000 0.0000
2451899.97215  2.63 1.79 2   0.00 0.00 0.00000 0.0000 0.0000
2452189.14851  1.79 2.37 2   0.00 0.00 0.00000 0.0000 0.0000
2452235.90051 -0.81 1.98 2   0.00 0.00 0.00000 0.0000 0.0000
2452538.06980  7.73 1.88 2   0.00 0.00 0.00000 0.0000 0.0000
2452651.88827 -7.96 2.36 2   0.00 0.00 0.00000 0.0000 0.0000
2452711.72343 -5.73 1.85 2   0.00 0.00 0.00000 0.0000 0.0000
2452712.74594  0.38 1.55 2   0.00 0.00 0.00000 0.0000 0.0000
2452988.87863  0.00 1.92 2   0.00 0.00 0.00000 0.0000 0.0000
2453339.97662 -1.19 1.85 2   0.00 0.00 0.00000 0.0000 0.0000
2453368.91795  0.24 1.96 2   0.00 0.00 0.00000 0.0000 0.0000
2453425.72302 -6.66 1.38 2   0.00 0.00 0.00000 0.0000 0.0000
2453425.72927 -7.53 1.88 2   0.00 0.00 0.00000 0.0000 0.0000
2453723.91920  1.93 1.74 2   0.00 0.00 0.00000 0.0000 0.0000
2453776.76985 -1.62 1.52 2   0.00 0.00 0.00000 0.0000 0.0000
2453776.77608 -2.64 1.50 2   0.00 0.00 0.00000 0.0000 0.0000
2453982.10746  4.57 1.30 2   0.00 0.00 0.00000 0.0000 0.0000
2454085.98215  3.44 2.05 2   0.00 0.00 0.00000 0.0000 0.0000
2454085.98875  2.72 2.06 2   0.00 0.00 0.00000 0.0000 0.0000
2454138.73882 -1.22 1.57 2   0.00 0.00 0.00000 0.0000 0.0000
2454398.04694 -1.97 1.66 2   0.00 0.00 0.00000 0.0000 0.0000
2454430.03169  0.71 1.99 2   0.00 0.00 0.00000 0.0000 0.0000
2454454.98751 -3.58 1.56 2   0.00 0.00 0.00000 0.0000 0.0000
2454455.85705 -1.14 1.31 2   0.00 0.00 0.00000 0.0000 0.0000
2454460.82264 -2.21 1.97 2   0.00 0.00 0.00000 0.0000 0.0000
2454718.13512  1.39 1.52 2   0.00 0.00 0.00000 0.0000 0.0000
2454719.13273  1.46 1.28 2   0.00 0.00 0.00000 0.0000 0.0000
2454723.12422  0.03 1.45 2   0.00 0.00 0.00000 0.0000 0.0000
2454724.13185 -0.17 1.57 2   0.00 0.00 0.00000 0.0000 0.0000
2454725.14330  1.29 1.68 2   0.00 0.00 0.00000 0.0000 0.0000
2455490.97157 -5.19 1.53 2   0.00 0.00 0.00000 0.0000 0.0000
2456695.57403 -0.31 0.64 3   0.00 0.00 0.00000 0.0000 0.0000
2456697.60659  0.21 0.54 3   0.00 0.00 0.00000 0.0000 0.0000
2456700.62164  0.00 0.53 3   0.00 0.00 0.00000 0.0000 0.0000
2456729.56735  1.40 0.54 3   0.00 0.00 0.00000 0.0000 0.0000
2456730.55791 -0.06 0.57 3   0.00 0.00 0.00000 0.0000 0.0000
2456731.52289 -1.26 0.84 3   0.00 0.00 0.00000 0.0000 0.0000
2456732.54351 -0.59 0.77 3   0.00 0.00 0.00000 0.0000 0.0000
2456735.53824  0.06 0.65 3   0.00 0.00 0.00000 0.0000 0.0000
\end{verbatim}

\label{lastpage}

\end{document}